\documentclass[12pt,preprint]{aastex}

\shorttitle{Effects of the South Atlantic Anomaly}
\shortauthors{Augusto et al.}

\begin{document}


\title{Effects of the South Atlantic Anomaly \\
    on the muon flux at sea level}


\author{C. R. A. Augusto, J. B. Dolival, C. E. Navia, and K. H. Tsui}
\affil{Instituto de F\'{\i}sica, Universidade Federal Fluminense, 24210-346,
Niter\'{o}i, RJ, Brazil}
\email{navia@if.uff.br}

%\and

%\author{R. J. Hanisch\altaffilmark{5}}
%\affil{Space Telescope Science Institute, Baltimore, MD 21218}


%\altaffiltext{1}{Visiting Astronomer, Cerro Tololo Inter-American Observatory.
%CTIO is operated by AURA, Inc.\ under contract to the National Science
%\altaffiltext{2}{Society of Fellows, Harvard University.}
%\altaffiltext{3}{present address: Center for Astrophysics,
%    60 Garden Street, Cambridge, MA 02138}
%\altaffiltext{4}{Visiting Programmer, Space Telescope Science Institute}
%\altaffiltext{5}{Patron, Alonso's Bar and Grill}


\begin{abstract}

The goal of this study is to examine the response and changes of the muon intensity at ground,   
due to magnetic anomaly over south Atlantic. Based on the data of two directional muon telescopes and located at 22S and 43W. These coordinates are inside of the South Atlantic Anomaly (SAA) region, a dip in the magnetosphere. 
This characteristic offers to the muon telescopes the lowest rigidity of response to cosmic protons and ions ( $\geq 0.4$ GV). 
The magnetosphere's dip is responsible for several processes, such as the high conductivity of the atmospheric layers due to the precipitation of energetic  particles in this region and an zonal electric field known as the pre-reversal electric field (PRE) with an enhancement at evening hours. In addition the open magnetosphere, propitiate the magnetic reconnections of
the IMF lines that will take place in this site in the day side. These factors are responsible for an unusually large particle flux present in the SAA region, including particles with energies above the pion production threshold.   
The main effect is an increase of  the muon intensity  ($E_{\mu}>0.2GeV$) at ground, in the day side, in up to ten times. We show that it is correlated with the pre-reversal electric field, and propitiate the observation of 
muon enhancements due to small solar transient events, such as corotating interaction region (CIR) and micro-flares. Details of these results are reported in this paper.

\end{abstract}





\section{Introduction}

The South Atlantic Anomaly (SAA) is the region where Earth's inner van Allen radiation belt makes its closest approach to the planet's surface. The Earth magnetic field  intensity is smaller over this rregion than elsewhere, caused by the fact that the center of Earth's magnetic field is offset from its geographic center by 280 miles \citep{frasier87}. This hole in the almost spherical magnetosphere
has been observed by spacecraft (Skylab, Mir) \citep{badhwar97}, and by Topex-Poseidon spacecraft as an enhancement of energetic particles in this region \citep{heirtzler02}. The presence of energetic protons above the 50 MeV level shows a 1000 times increase when compared to
other regions exterior to the SAA. The SAA has been confirmed by ground based experiments using a variety of instrument, especially observations of enhanced ionization induced by particle precipitation in this region \citep{abdu77,basu01}, the effect of such enhanced ionization in causing
enhanced ionospheric conductivity. The SAA covers a vast region
of the South Atlantic and its center is located at 26S and 53W and coincides with the Brazilian southern region. 
Fig.1 shows the geomagnetic total field intensity distribution, represented by iso-intensity lines over the globe.
The lowest value of magnetic intensity situated in South Brazil defines the position of the SAA region. 

On the other hand, since April 2007 we are operating the phase II of the Tupi experiment, with two directional muon telescopes at sea level and located at 22S and 43W, inside the SAA region and close to its central area. 
This characteristic supplies to the muon telescopes the lowest rigidity of response to cosmic protons and ions ( $\geq 0.4$ GV). This value is approximately half of the rigidity of the polar regions.
In addition, the SAA is like a hole in the magnetosphere and the IMF lines overtake the surface, in the day side, in this region. Consequently, the magnetic anomaly works like a funnel for incoming charged particles from space.
The main effect of the SAA is an increase of the muon intensity at sea level in this region, in up to 10 times in the day side.

The precipitation of energetic particles on the SAA ionizing the high layers of the atmosphere and increasing their conductivity  \citep{gledhill66,paulikas75,nishino02} and propitiating the appearance of an electric field, the pre-reversal electric field (PRE), with an enhancement at evening hours, the so called sunset enhancement.  
In fact, we show in this survey that there is a correlation in the hourly variation of the atmospheric conductivity gradient and the hourly variation of the muon intensity at ground, both present the so called sunset enhancement.  
In most cases the precipitation of the high energy particles in the SAA region begins around 3 hours after the sunrise and it finishes around one hour after the sunset, these schedules are subject to periodic variations (they move with the seasons of the year).

The low rigidity of response of the telescopes plus the fact that in the SAA region we have an almost open magnetosphere, and it propitiate the reconnection of the IMF lines, producing field lines with one end at the Earth and the other in distant space, are probably responsible for the observation of muon enhancements in association with solar transient event of small scale, such as the increase in the intensity of muons in coincidence with the gradual increase observed in the speed and density of the solar wind (initiate in solar holes), due to a corotating interaction region (CIR).  We have also observed two ground level enhancements (GLEs) in very close schedules of the sunset and in temporal coincidence with micro-flares whose X-ray prompt emission have been observed by GOES11. This means that the GLEs are associated with flares located near to the foot-point of the garden hose field line between the Sun and Earth.

We have also observed several sharp peaks in the muon counting rate, produced by cosmic ray precipitation in the SAA region, they are in temporal coincidence with the Swift-BAT triggers (a multi-band spacecraft Gamma Ray Burst detector)
and MILAGRO triggers (a ground based experiment located near to Los Alamos Laboratory). 
These triggers at least the observed in the Swift spacecraft are considered as noise triggers that occurred while Swift spacecraft was entering the SAA region \citep{augusto08}.

This paper is organized as follows: In Sec.II The Tupi experiment is described, in Sec.III the hourly and daily muon intensity variation in the SAA region, as observed by Tupi experiment, is described. 
The results on the muon enhancement at ground due to Corotating Interaction Region  and micro-flares are presented in Sec.IV and Sec.V respectively, and Sec.VI contains conclusions and remarks.

\section{The Tupi experiment}

Starting from April of 2007 we have initiated the Phase II of the Tupi experiment which consists of two identical muon telescopes built on the basis of plastic scintillators. The aim is a search for the daily muon intensity variations. One telescope has vertical orientation and the another has a 45 degree orientation with the vertical (zenith) pointing to the west. Both of them have an effective aperture of $65.6\;cm^2\;sr$. The Tupi experiment is at sea level and is located at 22S and 43W, which is close to the SAA center 26S and 53W. The telescopes are inside a building under two flagstones of concrete allowing registration of muons with $E_{\mu}\geq 0.2\;GeV$ required to penetrate the two flagstones, as shown in Fig.2. 

Both telescopes have a rigidity of response to cosmic proton (ions) spectrum above 0.4 GV, given by the local geomagnetic cutoff at 0.4 GV. The vertical telescope uses a veto or anti-coincidence guard with respect to the inclined telescope, and vice-verse, as is shown in Fig.3 where the logic is shown. Therefore, only muons with trajectories close to the telescope axis are registered. This guarantees the directionality of the muon telescopes. The data acquisition is made on the basis of the Advantech PCI-1711/73 card with an analogical to digital conversion rate of up to 100 kHz.

The primary cosmic ray particles (i.e. protons and nuclei) can be inferred through the detection of muons by telescopes at ground and underground levels.  The upper layers of the Earth's atmosphere is bombarded by a flux of cosmic primary particles. The chemical composition of this primary cosmic particles depends on the energy region. In the low energy region (above 10 GeV to several TeV), the dominant particles are protons ($\sim 80\%$). The primary cosmic rays collide with the nuclei of air molecules and produce an air shower of particle that include nucleons, charged and neutral pions, kaons etc. These secondary particles then undergo electromagnetic and nuclear interactions to produce yet additional particles in a cascade process. Of particular interest is the fate of charged pions, $\pi^{\pm}$, produced in the cascade. Some of these will interact with air molecule nuclei via the strong interaction, but others will spontaneously decay into a muon, $\mu^{\pm}$, plus a neutrino or anti-neutrino, $\nu_{\mu}$, following the scheme $\pi^{\pm}\rightarrow \mu^{\pm} \nu_{\mu}$ via the weak interaction.
Muons are quite penetrating and can reach the ground and enter the laboratory through the walls or roof of the building, and be detected with a suitable apparatus. Very high energy muons reach underground levels.

In order to estimate the counting rate of muons reaching detection level that are initiated by a flux of primary particles (protons),
the so called response function is obtained, and it is related as
\begin{equation}
\frac{dN(E,x,t)}{dE}=S(E,x)\frac{dJ(E,t)}{dE},
\end{equation}
where $dN(E,x,t)/dE$ is the counting rate of detector located at depth $x$, at the time $t$, for primary particles withing energy $E$ to $E+dE$ arriving withing a small solid angle near to vertical direction, $S(E,x)$ is the response function and $dJ(E,t)$ is the primary cosmic ray spectrum. Here, the response function is derived on the basis of the results obtained with the FLUKA Monte Carlo program to taken into account the muon production and propagation on the atmosphere \citep{fasso01,poirier02}.
The response function expressed as the number of muons at detection level (sea level) per incident proton primary is plotted in Fig.4 for two primary incident directions. 

\section{The hourly and daily muon intensity variations in the SAA region} 

\subsection{The hourly variations}

We show in this survey that the muon intensity at ground (muons with energies above 0.2 GeV)  in the SAA region increases in up to ten times in the day side. The effect is due to the precipitation of energetic (KeV to MeV) particles,  the trapped and azimuthally drifting energetic
particles, bouncing between hemispheres, come deeper down into the atmosphere owing to the low field
intensity over SAA, thereby interacting with the dense atmosphere resulting in ionization production and increasing their conductivity. 
In addition, the open magnetosphere (magnetosphere's hole), propitiates the magnetic reconnections of
the IMF lines that will take place in this site in the day side. There is also an enhanced zonal electric field known as the pre-reversal electric field enhancement (PRE) \citep{abdu05}. The PRE arises from two combined effects, the eastward thermospheric wind and the longitudinal conductivity gradient. These factors are responsible for an unusually large particle flux present in the SAA region and that include particles (protons-ions) with energies above the pion (muon) production energy threshold.

In fact, we show that there is a correlation in the hourly variations of the PRE and the hourly variations of the muon intensity at ground, both present the so called sunset enhancement. In the upper panel of Fig.5, the PRE obtained by simulation under several conductivity gradients is shown, and in the lower panel, the two month (April-May 2007 )averages of the hourly variations of the muon intensity is shown. It is interesting to observe that
the muon intensity in the vertical telescope is almost twice than the muon intensity in the inclined telescope, in agreement with the expected zenith angle dependence for the muon intensity, and related as $\cos(\theta)^2=0.5$ for $\theta=45^0$ in the $\sim$GeV region.

On January 24, 2008, a new and faster micro-computer was used for the data acquisition system of the muon telescopes, and the result was an increment in the counting rate of at least 10 times. Under this new condition, it was possible to observer the fine structure of the muon counting rate (raw data) showing details of the precipitation of energetic particles in the SAA region. Fig.6 summarizes the situation, in the upper panel
the muon counting rate at every 10 seconds (raw data) is shown, where the sharp peaks are due to the precipitation of high energy cosmic ray in the SAA region and producing muons in the atmosphere. In the lower panel, the counting rate at every 5 minutes is shown for comparison.
This results is at January 28, 2008, (summer in the Souther hemisphere). In both panels, it is possible to see the muon enhancement at evening hours due to pre-reversal electric field enhancement (PRE).

The precipitation of high energy particles in the SAA region is subject to fluctuations. For instance there are days where no precipitation of particles in this region is observed. An example is shown in the Fig.7, where the muon counting rate at every 5 minutes and for 5 serial days
is shown. It is possible to verify that in the days 4 and 5 of May, there is no precipitation of particles. The origin of these fluctuations
can be described using the ``Open Magnetosphere'' Models  \citep{dungey61,debrito05} based on a "reconnection process" that takes place at the front (dayside) of the Magnetosphere to produce field lines with one end at the Earth and the other in distant space.
The magnetic reconnections of the Interplanetary Magnetic Field (IMF) take place in the SAA region. However, the coupling and  specially the decoupling process under which the IMF evolves and carries out the task of transporting particles out of the SAA region is not still very well understood . In short, in most cases the precipitation the high energy particles in the SAA region begins around 3 hours after the sunrise and finishes around one hour after the sunset. 

\subsection{The daily variations}

The solar daily variations known also as the diurnal solar anisotropy of the cosmic ray intensity has been observed by ground and underground based detectors, covering a wide range of the cosmic ray spectrum, with rigidities between 10 GV to 400 GV. The anisotropy reflects the local interplanetary cosmic ray distribution, and it is widely believed that the anisotropy arises when the  cosmic ray particles co-rotate with solar wind stream, following the IMF lines, and it is related in terms of diffusion convection and drift of  cosmic ray in the IMF.

According to \citep{forman75}, at 1 AU the co-rotating stream (solar wind) has a speed of order 450 $km\;s^{-1}$ (in average) and at $\sim 18h$ (local time) approximately in the same direction as the Earth's orbital motion (of 30 $km\;s^{-1}$). In other words, the co-rotating stream will overtake the Earth ``almost in the vertical'' from the direction of $\sim 18\;h$, and it is known as the phase of the anisotropy. This phase has shifted toward earlier hours $\sim 15h$ during the lower solar cycle due to drift process. So far, the drift process is still the most likely and accepted explanation of the phase shift. However, the solar daily variations specially observed in the high rigidity region, have shown remarkable changes in phase and amplitude during long periods of observation.  A review of characteristics of the observed cosmic ray diurnal variations over three decades has been reported by
\citep{Ahluwalia97,mori96}.  

In the SAA region the cosmic ray daily variation is amplified. We have used the fast Fourier transformation (FFT) for studying periodicities and scaling properties that might be present in our time series constructed using the hourly muon intensities. The power spectra of the hourly muon intensities for two months are show in Fig.8. In this case, there is a series of peaks, such as at 0.99 days (daily anisotropy), and a small peak can be also observed at 0.58 days known as the semi-diurnal anisotropy.
There is also a strong signal of the harmonic $27/n$, with $n=4$ giving a peak at $\sim 7$ days, which corresponds to the quasi-periodic corotating streams that occurs due to solar rotation period of 27 days.  We would pointed out that the harmonics like $27/n$ has been observed in the power spectra of solar wind speed measurements reported by Burlaga and Lazarus \citep{burlaga00}. These results strongly suggest that the solar wind and the protons (ions), producing the muon intensity, are modulated by the Sun.

\section{Muon enhancement at ground due to Corotating Interaction Region}

Corotating interaction regions (CIRs) are regions of compressed plasma, formed
at the leading edges of corotating high-speed solar wind streams, originating in
coronal holes as they interact with the preceding slow solar wind. They are particularly
prominent features of the solar wind during the declining and minimum
phases of the 11-year solar cycle. However, in most of cases, the particle enhancement observed at ground due to a CIR occurs at times of higher solar activity, interspersed with slow solar wind and transient flows associated with
coronal mass ejections (CMEs) \citep{iucci79,richardson02}.

Although the corotating shocks that
often develop at the boundaries of CIRs beyond 1 AU appear to play a role in
the acceleration of interplanetary ions including  cosmic rays. The CIRs with developed forward and reverse shocks are occasionally observed at 1 AU. An example of corotating high-speed solar wind streams originating in
coronal holes and observed in October 2007 by ACE spacecraft is shown in Fig.9 including their associated muon (Tupi) enhancement. The muon enhancement associated with the forward shock is clearly evident. In addition, the EPAM detector
on board of the ACE spacecraft had also observed proton (ions) enhancement, exactly in coincidence with the beginning of the forward shock, consequently in coincidence with the beginning of the muon enhancement. Fig.10 summarizes the situation, where  the barometric pressure is also included. It can be noticed that the change (increase) in the muon counting rate happens during the increase of the barometric pressure, as the flow of muons is anti-correlated with the pressure, the change in the muon flow is not due to pressure variations.

\section{Muon enhancements at ground in association with solar micro-flares}

A solar flare is defined as a sudden, rapid, and intense variation in brightness. A solar flare occurs when magnetic energy that has built up in the solar atmosphere is suddenly released. Radiation is emitted across virtually the entire electromagnetic spectrum. The first solar flare recorded in astronomical literature was on September 1, 1859. Two scientists, Richard C. Carrington  and Richard Hodgson \citep{carrington60}, were independently observing sunspots at the time, when they viewed a large flare in white light. In solar flares, the interaction of the energetic electrons with thermal protons provides the deceleration, and
X-ray photons with energies less than or nearly equal to the electron energy are produced. These X-ray photons are the emitted radiation signatures detected by scientific instruments, such as GOES and SOHO.
The frequency of flares coincides with the eleven year solar cycle. When the solar cycle is at a minimum, active regions are small and rare and few solar flares are detected. The occurance increases in number as the Sun approaches the maximum of its cycle. However, the period around the solar minimum is useful to observe small transient events, such as the micro flares.

Micro-flares (flares releasing $10^{27}-10^{28} ergs$) have been identified as changes in coronal emission due to impulsive heating of new materials to coronal temperature \citep{krucker98,berghmans98,aschwand00}. The measurements of these micro events are made possible by SOHO/EIT and TRACE. The soft X-ray flares classified by GOES as class A ($1.0-9.9\times 10^{-8}Watts\;m^{-2}$) are usually associated with ``micro-flares''.
The soft X-ray emission for solar flares is a signature of acceleration of electrons. Are there accelerated ions in these processes as well? The answers is yes, while their fraction and chemical composition as well as the relationship between flares and coronal heating are still poorly understood. Most of the accelerated particles should contribute to heating, and probably only a small fraction of them escapes to space. In addition, it has been claimed that the micro-events have many physical properties in common with regular flares, while their locations and statistical behaviors are different.

Ground-level solar cosmic ray events are usually observed by high latitude neutron monitors at relatively low rigidities ($\sim 1-3$, GV) and in most cases the ground-level events are linked to solar flares of high intensity
whose prompt X-ray emission is cataloged as X-class (above $10^4Wm^{-2}$). Evidently, the solar flare detection at ground depends on several aspects, such as a good magnetic connection between the Sun and Earth.
We show in this paper that the micro-flares, or at least the bigger ones of them, are capable of accelerating ions up to energies beyond 100 GeVs in the tail of the energy spectrum.
We reporter here, the association of two muon enhancements with two micro flare events, whose X-ray pronth emission
have been reported by GOES11.

(a) The first event was observed at 22.98 UT on April 24, 2007. 
Its (0.5-4.0 \AA) soft X-ray peak is a 2.7B class, and its (1.0-8.0 \AA) soft X-ray peak is a 5.9C class. 
The GOES soft-X ray light curves of the events are shown in Fig11 (upper panel), together with their associated Tupi muon enhancements (lower panel) having a confidence level of $9.1\sigma$.

(b) The second event was observed at 19.67 UT on June 12, 2007. 
Its (0.5-4.0 \AA) soft X-ray peak is a 1.9A class, and its (1.0-8.0 \AA) soft X-ray peak is a 5.9B class. 
The GOES soft-X ray light curves of the events are shown in Fig.12 (upper panel), together with their associated Tupi muon enhancements (lower panel) having a confidence level of $7.5\sigma$.

In both cases, the ground level enhancements (GLE) peaks are advanced in relation to the (1.0-8.0 \AA) X-rays peaks (21 minutes and 12 minutes respectively), this means that in these two cases the high energy protons (ions) are firstly emitted. The same behavior is observed in the pronth X-ray emission, there is an advance of the 
(0.5-4.0 \AA) X-ray peak (more energetic) in relation of  the (1.0-8.0 \AA) soft X-ray peak. 
Most solar flares associated with GLEs are located on the western sector of the Sun where the IMF is well connected to the Earth. 
The two GLEs here analyzed are associated with flares located near to the foot-point of the garden hose field line, because the protons (ions) travel toward the Earth following the garden hose line.  Fig.13 summarizes the situation, for the best condition, pitch angle, $\theta =45^0$,  where the pitch angle defined as the angle between the sun-ward direction and the telescope axis direction. In this case, usually the protons arrive promptly and have very sharp onsets.
Conversely, GLEs associated with flares far from the garden hose field line are usually delayed in their arrival at Earth and have more gradual increases to maximum intensity. 

The pitch angles for these two events when they reach the maximum intensity as observed by the two telescopes are represented in Fig.14. The best condition (pitch angle $\sim 45^0$) is satisfied by the inclined telescope. These two examples, has been included in our collection on muon enhancement in association with solar flares of small scale \citep{navia05,augusto05}.

\section{Conclusions}

The Earth is surrounded by an almost spherical magnetic field, the magnetosphere, and it is a
natural shielding of the Earth surface to solar particle  and
cosmic rays up to several GeV energies. Specially in the day side, the supersonic solar wind interaction with the magnetosphere produces a bow shock, shielding the Earth of several types of radiations (charged particles) coming from external space.  
However, at a certain location over the South Atlantic Ocean, and centered in the south of Brazil, the shielding effect of the magnetosphere is not quite spherical but shows a hole, as a result of the eccentric displacement of the center of the magnetic field from the geographical center of the Earth. This is the so called SAA region. This behavior of the magnetosphere is responsible for several processes, such as the high conductivity of the atmospheric layers due to the precipitation of energetic  particles in this region. The gradient of this high conductivity is in part responsible for an zonal electric field known as the pre-reversal electric field enhancement (PRE) with an enhancement at evening hours, the so called sunset enhancement.
In addition the open magnetosphere of the SAA region propitiates the magnetic reconnections of
the IMF lines that will take place in this site in the day side. These factors are responsible for an unusually large particle flux present in the SAA region.

We have reported the effects of the SAA in the muon flux at sea level, the data comes from two direction muon telescopes  (Tupi experiment) and located inside the SAA region. This characteristic supplies to the telescopes the lowest rigidity of response ($>0.4$ GV) to solar and cosmic protons (ions). The main results are summarized as: 

(a) The precipitation of very high energy particles increases in up to ten times (see Fig.7) the muon intensity  ($E_{\mu}>0.2GeV$) at ground. In most cases, the precipitation in the SAA region begins around 3 hours after the sunrise and  finishes around one hour after the sunset. These schedules are subject to periodic variations. 

(b) We have found a direct correlation between the hourly variation of the PRE  and the hourly variation of the muon flux, not only their monthly averages, but also the hourly variation in an only day  (see Fig.5 and Fig.6). In all cases is possible to see, the so called sunset enhancement.

c) The observation of the precipitation of high energy particles in association with Swift-BAT and MILAGRO triggers. The Swift-BT triggers  occurred while Swift spacecraft was entering the SAA region. In addition, the MILAGRO triggers occurs for events whose coordinates point to the SAA region \citep{augusto08}.

(d) The SAA region propitiates the observation of muon enhancements at ground due to CIR events at times the lower solar activity (see Fig.9 and Fig.10). In most cases, the particle enhancement observed at ground due to a CIR occur at times of higher solar activity, interspersed with slow solar wind and transient flows associated with coronal mass ejections (CMEs) \citep{iucci79,richardson02}. However, at times of lower solar activity, the CIRs are mainly associated to fast solar wind, originated in coronal holes, and they are observed mainly by spacecraft.

(e) The SAA region also propitiates  the observation of muon enhancement at ground due to solar flares of small scale, ``micro flares''. From the two events here reported, it is possible to deduce that the micro-events have many physical properties in common with regular flares. For instance, the tail of the particle energy spectrum in all cases
reaches many dozens of GeV energies, because they produce muons ($E_{\mu}>0.2$ GeV) in the Earth's atmosphere.

The Tupi experiment is in progress, and in a period of 6 months we will increase the number of telescopes of the current ones 2 to 14, increasing in 12 times  the field of view of Tupi experiment. Our objective is to take advantage of our location inside the anomaly to detect gamma ray burst in the range of GeV to TeV, because there are theoretical models that predict GRBs in this energy region. Clear, we will continue studying solar transient phenomenons and to explore other possible phenomena that happen inside and due to the anomaly.

\acknowledgments

This work is supported by the National Council for Research (CNPq) in Brazil, under Grants No. $479813/2004-3$ and $476498/2007-4$. 
We are grateful to the various catalogs available  on the web and to their open data policy, especially to the Space Environment Center (SEC), the ACE Real-Time Solar Wind (RTSW) Data and the GCN report.
For comment and suggest, please write to
\email{navia@if.uff.br}.

\clearpage

%% the file, uncomment the \plotone, \plottwo, and
%% If you need a layout that cannot be achieved with \plotone or
%% \plottwo, you can invoke the graphicx package directly with the
%% \includegraphics command or use \plotfiddle. For more information,
%% please see the tutorial on "Using Electronic Art with AASTeX" in the
%% documentation section at the AASTeX Web site,
%% http://www.journals.uchicago.edu/AAS/AASTeX.
%%
%% The examples below also include sample markup for submission of
%% supplemental electronic materials. As always, be sure to check
%% the instructions to authors for the journal you are submitting to
%% for specific submissions guidelines as they vary from

%% This example uses \plotone to include an EPS file scaled to
%% 80% of its natural size with \epsscale. Its caption
%% has been written to indicate that additional figure parts will be
%% available in the electronic journal.

\begin{figure}
\vspace*{-3.0cm}
\hspace*{-3.00cm}
\epsscale{1.30}
\plotone{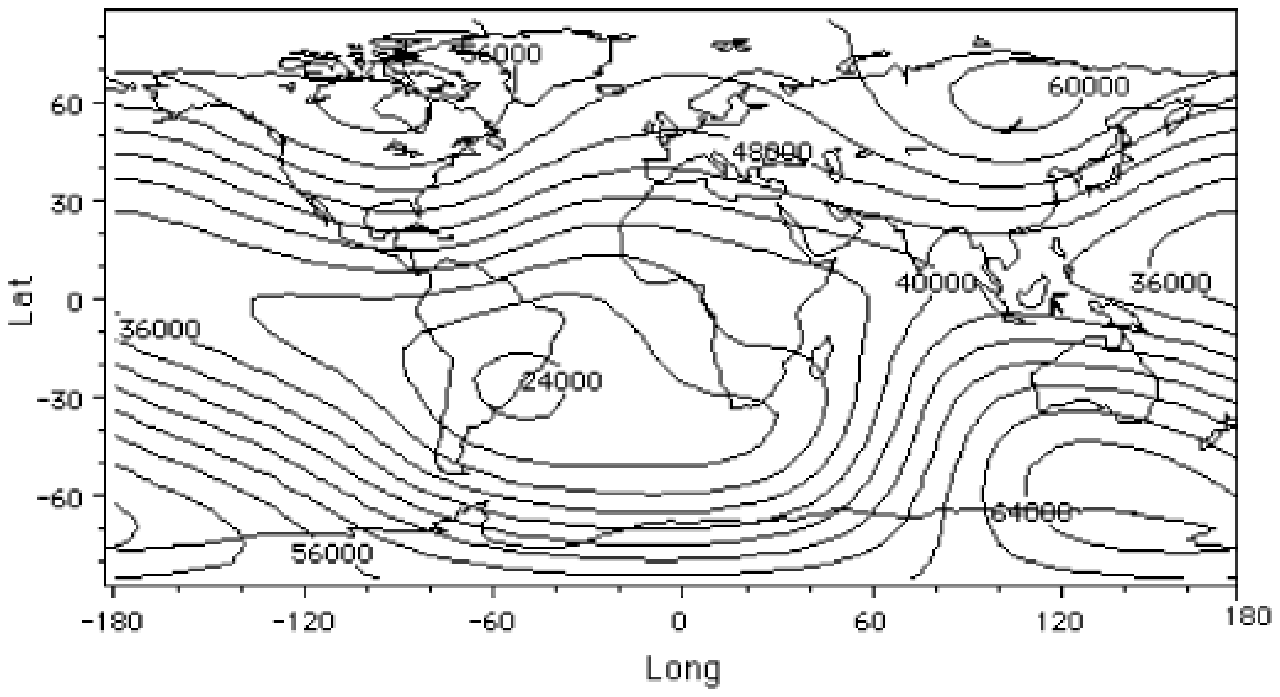}
\vspace*{-21.0cm}
\caption{The geomagnetic total field intensity distribution, represented by iso-intensity lines over the globe.
The lowest value of magnetic intensity situated in South Brazil define the position of the SAA region. 
}
\label{fig1}
\end{figure}

\begin{figure}
\vspace*{-2.0cm}
\hspace*{-1.00cm}
\epsscale{0.70}
\plotone{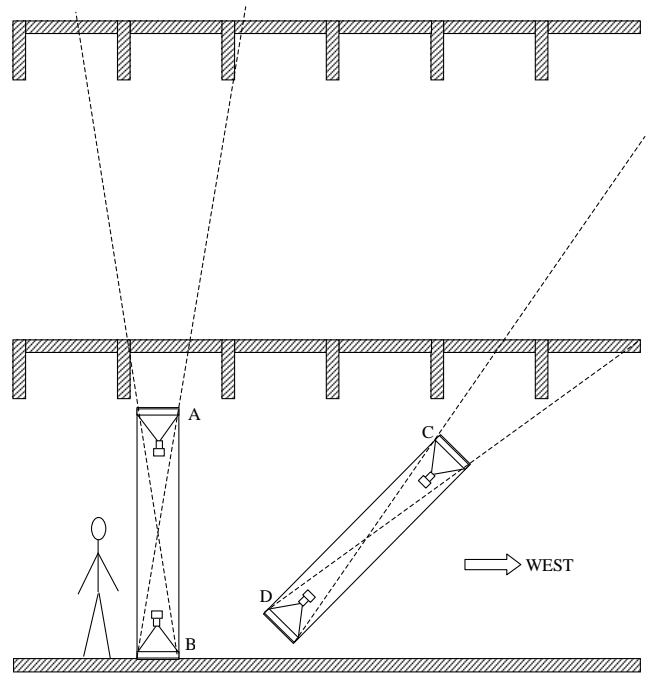}
\vspace*{-6.0cm}
\caption{Experimental setup of the Tupi experiment Phase II, showing the two telescopes.}
\label{fig2}
\end{figure}

\newpage
\clearpage

\begin{figure}
\vspace*{-5.0cm}
\hspace*{-0.00cm}
\epsscale{1.20}
\plotone{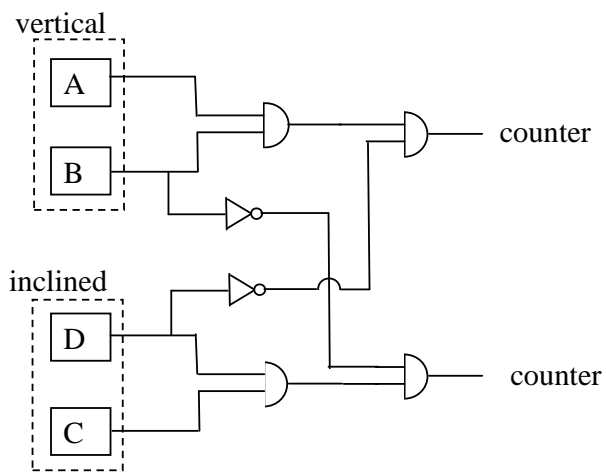}
\vspace*{-19.0cm}
\caption{General layout of the logic implemented in the data acquisition system. The vertical telescope use 
 a veto or anti-coincidence guard system with a detector of the inclined telescope and vice-verse. This system allow only the detection of muons traveling close to the telescope axis directions.}
\label{fig3}
\end{figure}

\begin{figure}
\vspace*{-2.0cm}
\hspace*{-0.00cm}
\epsscale{0.70}
\plotone{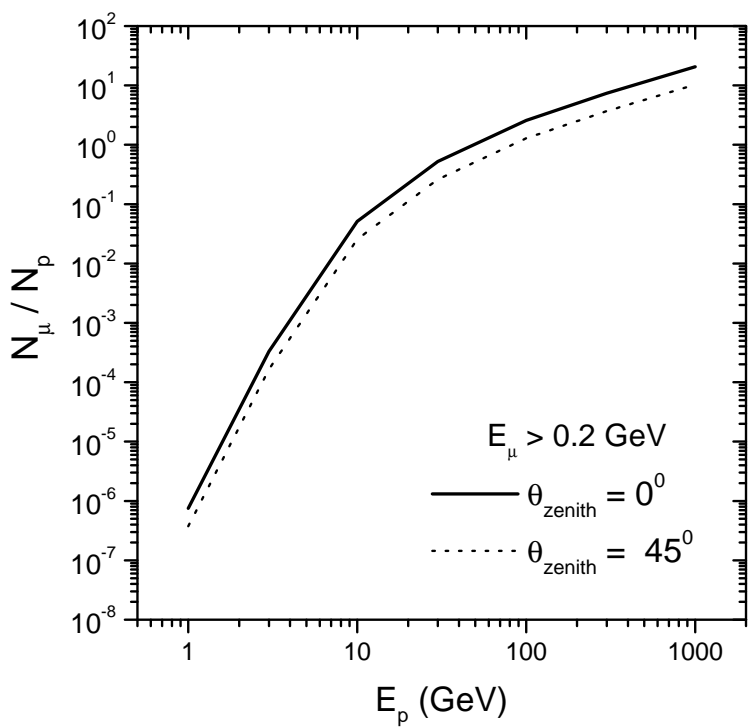}
\vspace*{-5.0cm}
\caption{Monte Carlo results with the FLUKA code, showing the number of muons which reach ground level per incident proton primary,
 as a function of the proton energy.}
\label{fig4}
\end{figure}

\newpage
\clearpage

\begin{figure}
\vspace*{-1.0cm}
\hspace*{-2.00cm}
\epsscale{1.40}
\plotone{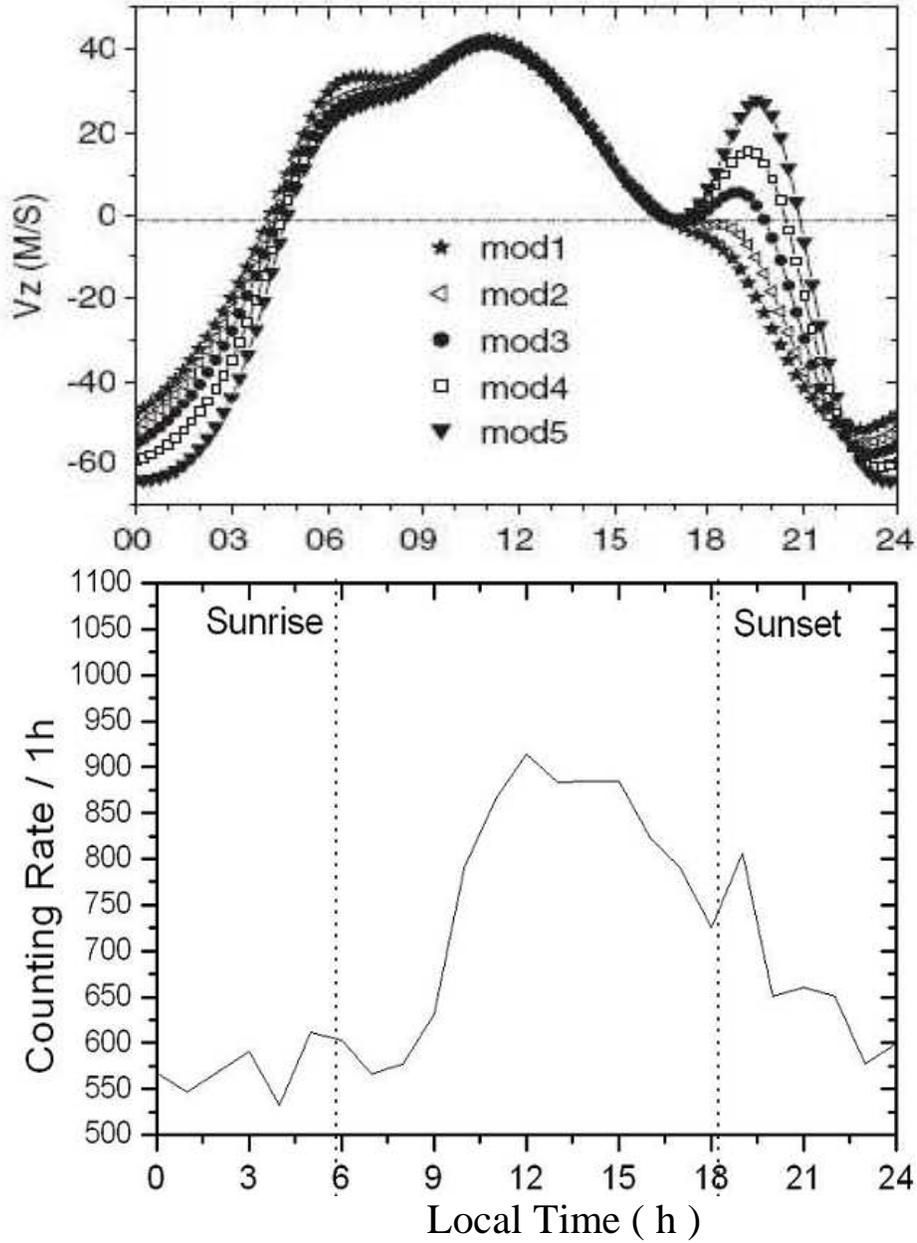}
\vspace*{-14.0cm}
\caption{Upper panel, Monte Carlo results of the hourly variation of the atmospheric pre-reversal electric field (PRE). The curves identified as mod 1 and mod 5 correspond to the lowest and highest conductivity gradients at sunset \citep{abdu05}. Lower panel, hourly variation of the muon intensity  observed by Tupi vertical telescope on February, 2008.}
\label{fig5}
\end{figure}

\newpage
\clearpage

\begin{figure}
\vspace*{-1.0cm}
\hspace*{-0.00cm}
\epsscale{0.80}
\plotone{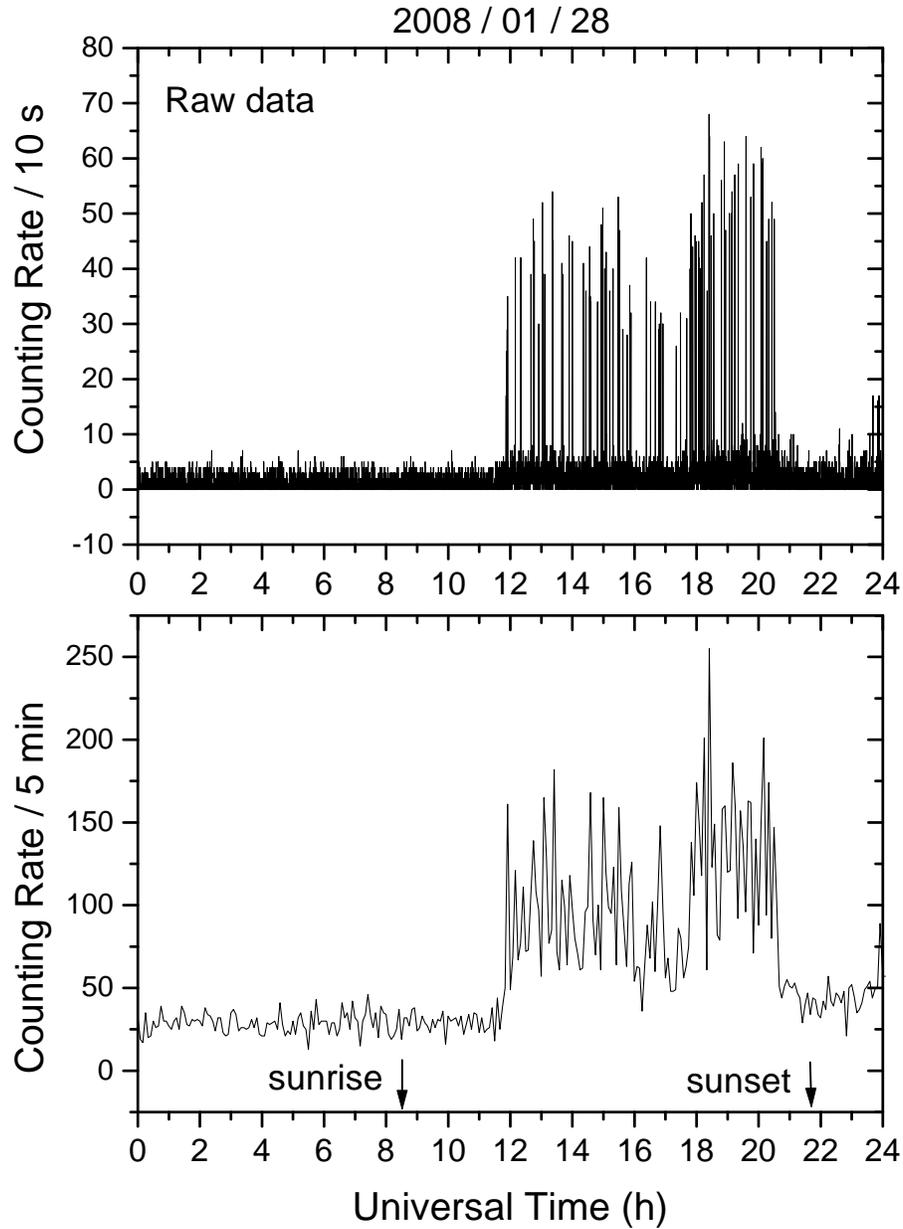}
\vspace*{-1.0cm}
\caption{Upper panel, the muon counting rate at every 10 seconds (raw data) and in the lower panel the muon counting rate at every five minutes, registered by the muon vertical telescope on January 28, 2008.}
\label{fig6}
\end{figure}

\newpage
\clearpage

\begin{figure}
\vspace*{-4.0cm}
\hspace*{-1.00cm}
\epsscale{1.10}
\plotone{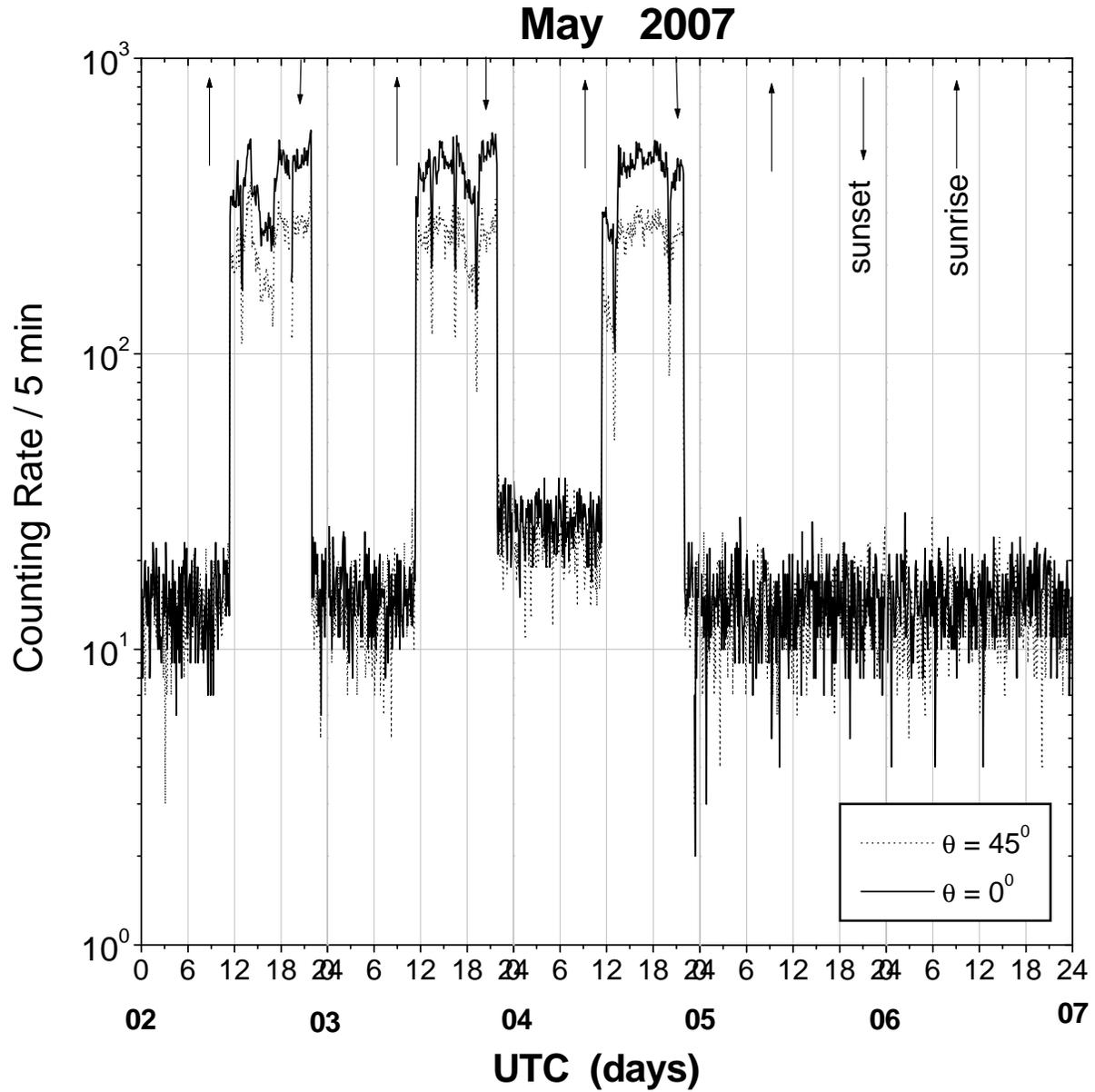}
\vspace*{-8.0cm}
\caption{The muon counting rate at every five minutes, for five consecutive days.
The bold and thin lines are the counting rate registered in the vertical direction and inclined direction ($45^0$) respectively. In all cases the muon energy threshold is $E_{\mu}\geq 0.2\;GeV$.}
\label{fig7}
\end{figure}

\newpage
\clearpage

\begin{figure}
\vspace*{-6.0cm}
\hspace*{-0.00cm}
\epsscale{0.80}
\plotone{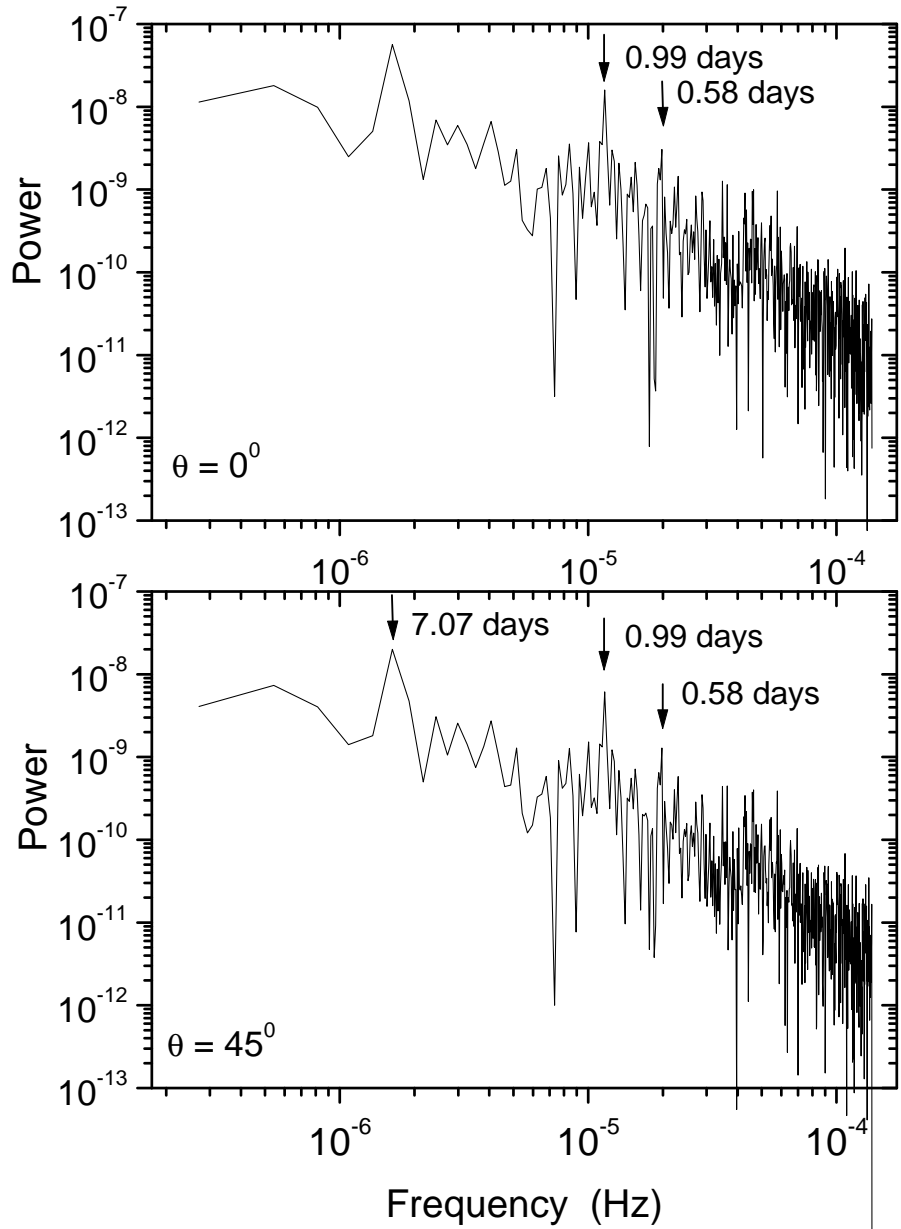}
\vspace*{-1.0cm}
\caption{Tho month power spectral density as a function of the frequency, obtained using the hourly averages of  the muon intensity and measured during two moths.}
\label{fig8}
\end{figure}

\newpage
\clearpage

\begin{figure}
\vspace*{-4.0cm}
\hspace*{-1.00cm}
\epsscale{0.70}
\plotone{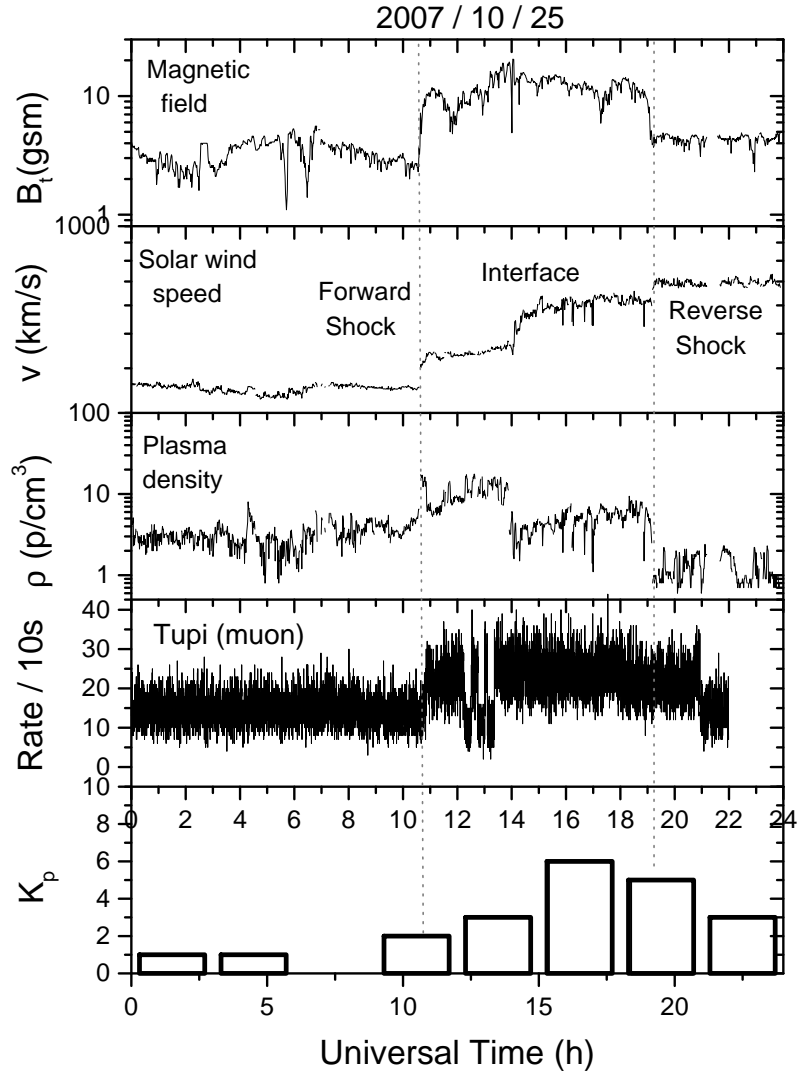}
\vspace*{-1.0cm}
\caption{Example of a CIR at 1 AU due to  corotating high-speed solar wind streams originating in
coronal holes with developed forward and reverse shock pairs and associated
particle (muon) enhancements, observed by Tupi telescope at sea level in October 25 2007. The solar wind parameters
have been observed by ACE solar wind spacecraft. The Kp index was obtained by ground based observations.}
\label{fig9}
\end{figure}
\newpage
\clearpage

\begin{figure}
\vspace*{-4.0cm}
\hspace*{-1.00cm}
\epsscale{0.70}
\plotone{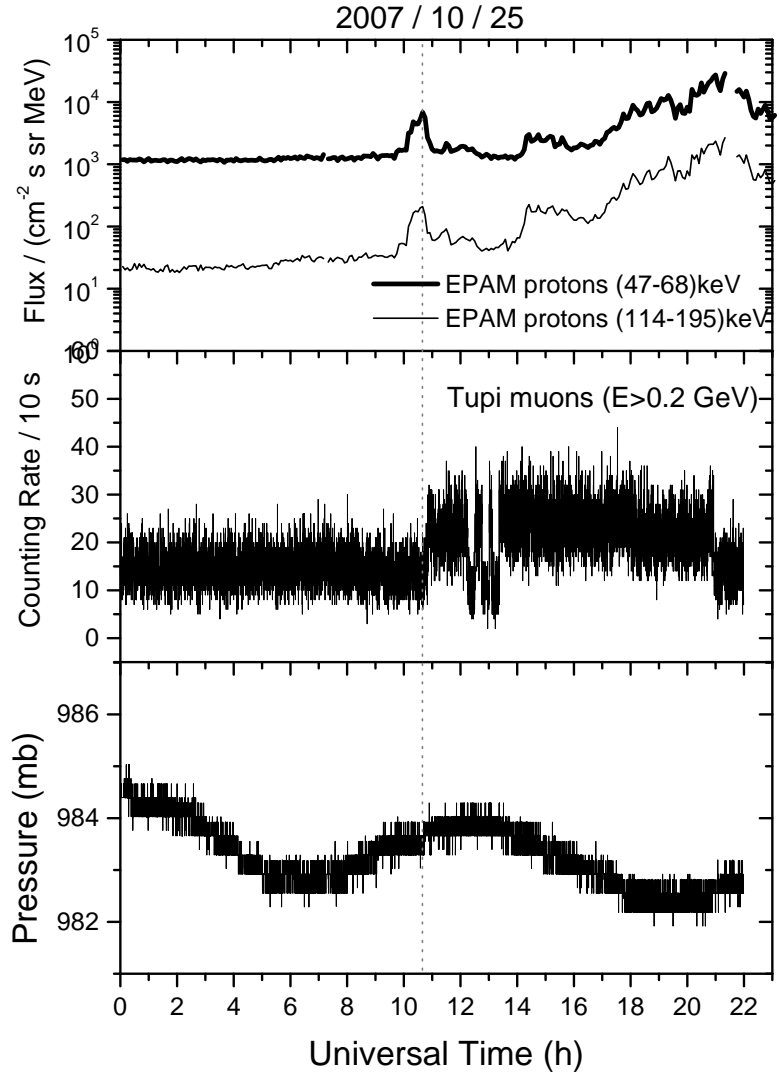}
\vspace*{-1.0cm}
\caption{The hourly variation: Upper panel, ACE-EPAM protons in two energy band. Middle panel The Tupi muon counting rate (raw data) and Lower panel the barometric pressure, measured at Tupi room, all on October 10, 2007.}
\label{fig10}
\end{figure}

\newpage
\clearpage

\begin{figure}
\vspace*{-6.0cm}
\hspace*{-1.00cm}
\epsscale{0.70}
\plotone{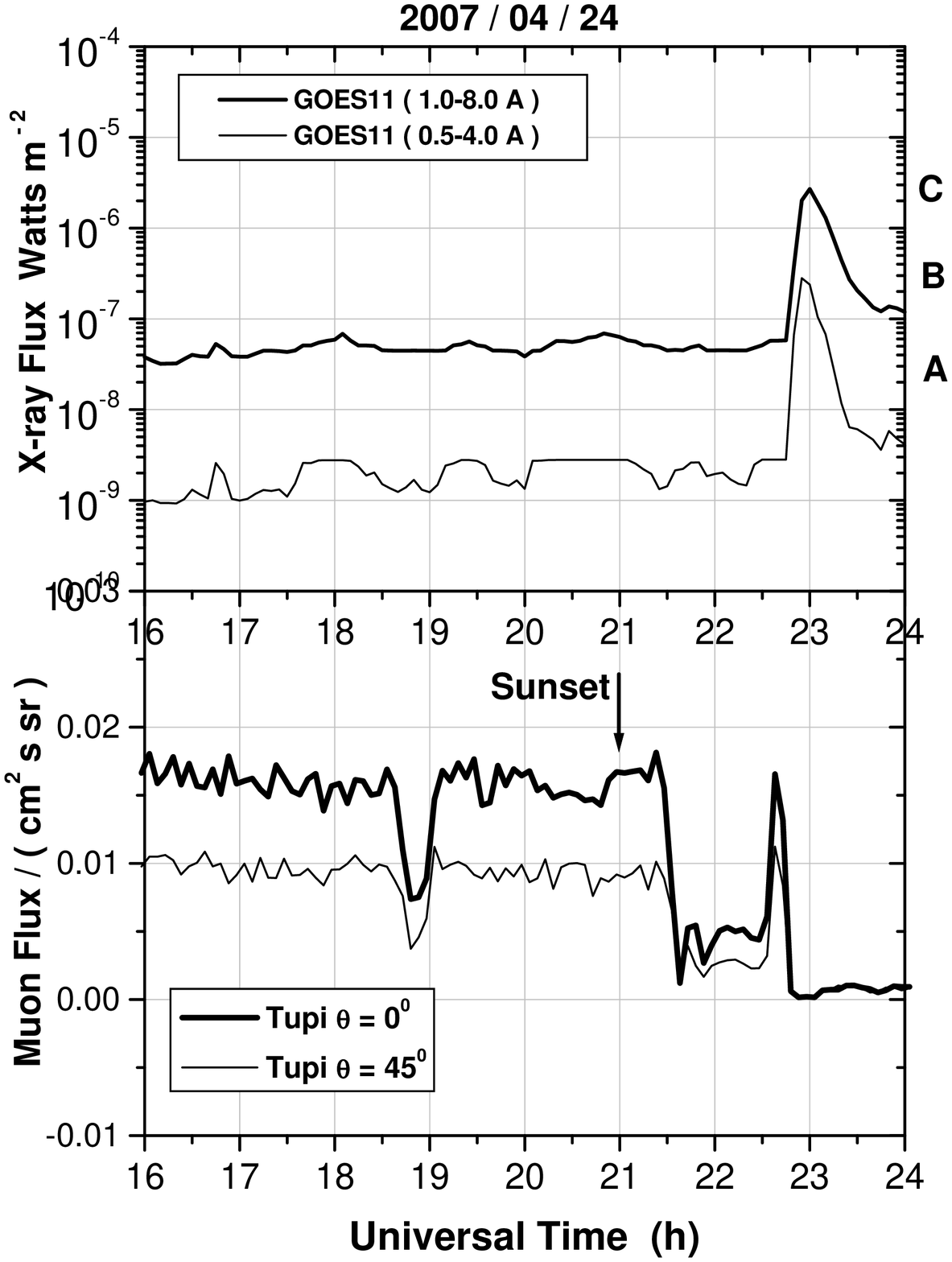}
\vspace*{-1.0cm}
\caption{Upper panel: The X-ray flux on April 24 2007 according to GOES11 for two wave length bands. Lower panel:
The 5 minutes Tupi integral muon intensities where the bold and thin lines are for the vertical and inclined directions
respectively.}
\label{fig11}
\end{figure}

\newpage
\clearpage

\begin{figure}
\vspace*{-6.0cm}
\hspace*{-1.00cm}
\epsscale{0.70}
\plotone{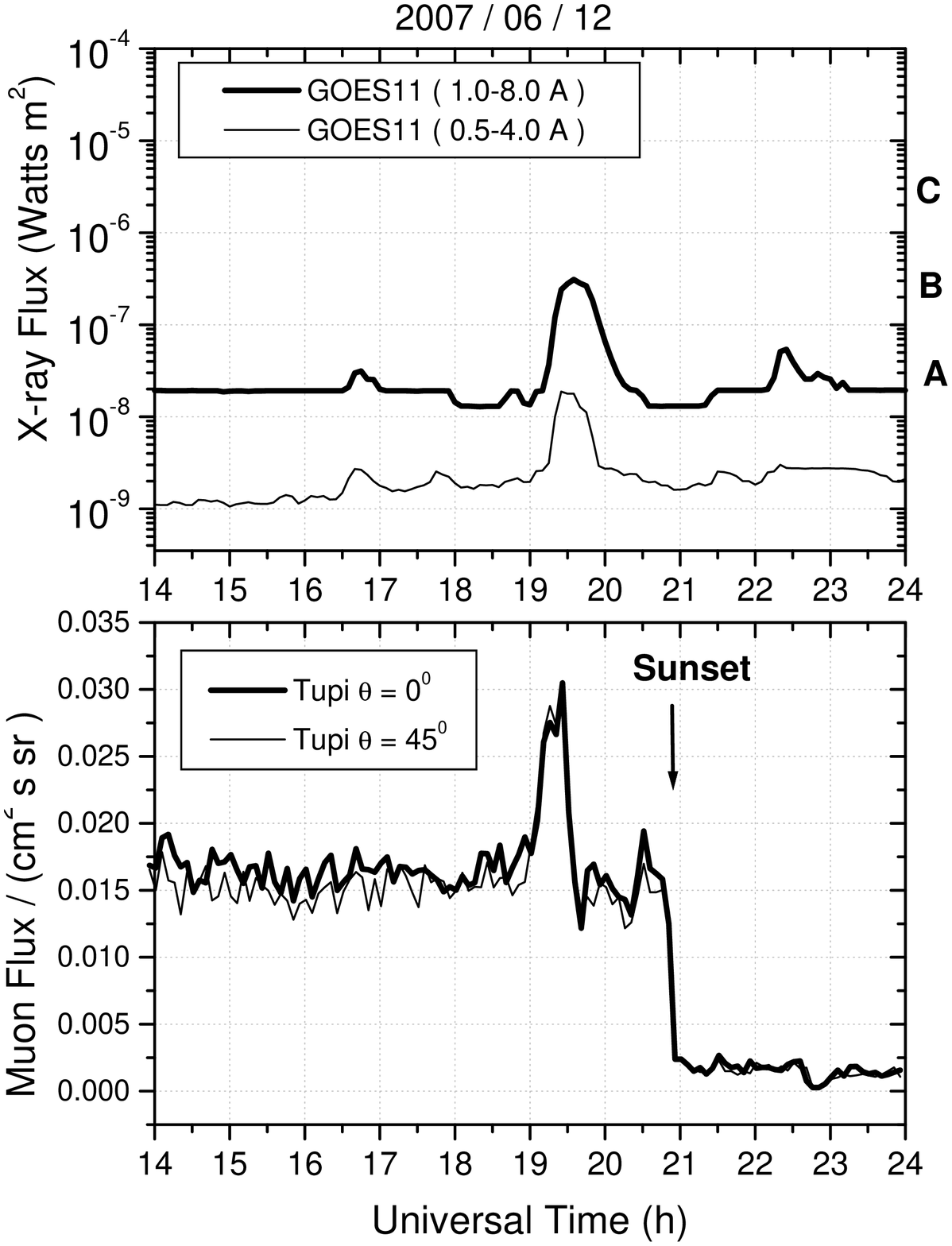}
\vspace*{-1.0cm}
\caption{Upper panel: The X-ray flux on Jun 12 2007 according to GOES11 for two wave length bands. Lower panel:
The 5 minutes Tupi integral muon intensities where the bold and thin lines are for the vertical and inclined directions
respectively.}
\label{fig12}
\end{figure}
\newpage
\clearpage

\begin{figure}
\vspace*{-4.0cm}
\hspace*{-1.00cm}
\epsscale{1.20}
\plotone{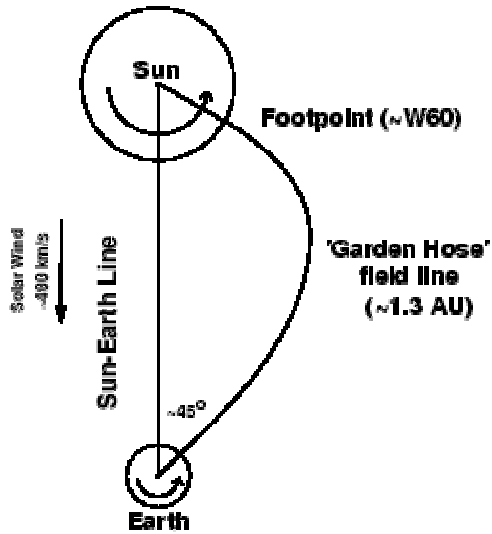}
\vspace*{-19.0cm}
\caption{Cartoon showing a good magnetic field connection between the Sun and Earth. Flares located near to the foot-point of the ``garden hose'' field line between the Sun and Earth, reach the Earth with a pitch angle close to $45^0$. Because, protons (ions) travel toward the Earth in a spiral trajectory, following the garden hose line and have very sharp onsets.
}
\label{fig13}
\end{figure}

\begin{figure}
\vspace*{-2.0cm}
\hspace*{-1.00cm}
\epsscale{0.70}
\plotone{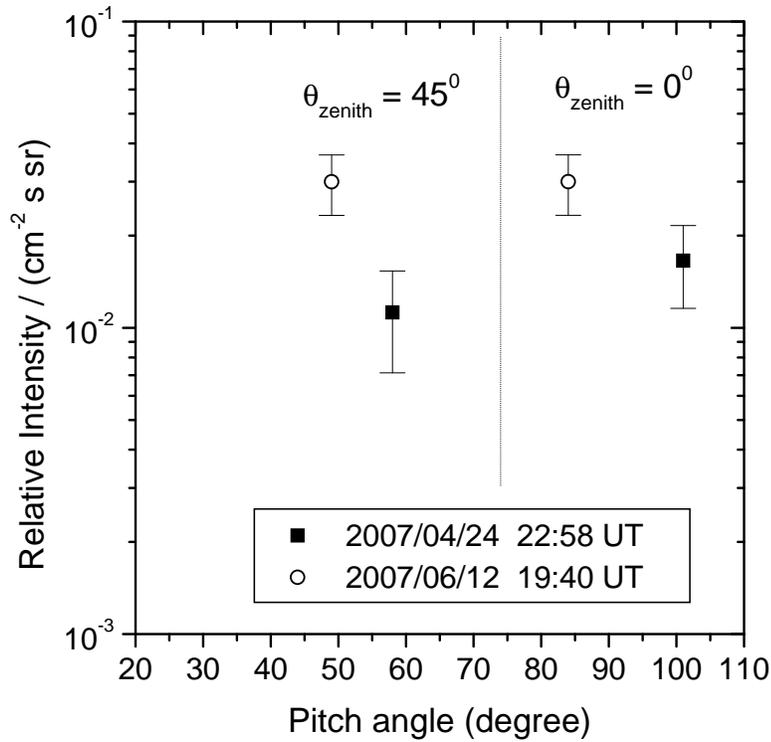}
\vspace*{-5.0cm}
\caption{The pitch angles for the two events (muon enhancements) associated with micro-flares, when they reach the maximum intensity as observed by the two telescopes.}
\label{fig14}
\end{figure}

%% Here we use \plottwo to present two versions of the same figure,
%% one in black and white for print the other in RGB color
%% for online presentation. Note that the caption indicates
%% that a color version of the figure will be available online.
%%

%% If you are not including electonic art with your submission, you may
%% mark up your captions using the \figcaption command. See the
%% No more than seven \figcaption commands are allowed per page,
%% so if you have more than seven captions, insert a \clearpage

%% Tables should be submitted one per page, so put a \clearpage before

%% Two options are available to the author for producing tables:  the
%% deluxetable environment provided by the AASTeX package or the LaTeX
%% table environment.  Use of deluxetable is preferred.
%%

%% Three table samples follow, two marked up in the deluxetable environment,
%% one marked up as a LaTeX table.

%% In this first example, note that the \tabletypesize{}
%% command has been used to reduce the font size of the table.
%% We also use the \rotate command to rotate the table to
%% landscape orientation since it is very wide even at the
%% Note also that the \label command needs to be placed

%% This table also includes a table comment indicating that the full
%% version will be available in machine-readable format in the electronic

\clearpage

\end{document}